\documentclass[sigconf]{acmart}

\AtBeginDocument{%
  }



\copyrightyear{2024}
\acmYear{2024}
\setcopyright{rightsretained}
\acmConference[HPDC '24]{The 33rd International Symposium on High-Performance Parallel and Distributed Computing}{June 3--7, 2024}{Pisa, Italy}
\acmBooktitle{The 33rd International Symposium on High-Performance Parallel and Distributed Computing (HPDC '24), June 3--7, 2024, Pisa, Italy}\acmDOI{10.1145/3625549.3658688}
\acmISBN{979-8-4007-0413-0/24/06}

\usepackage{algorithm}
\usepackage[noend]{algpseudocode}
\usepackage{subcaption}
\usepackage{multirow}

\newcommand{\projectname}{Loki}

\usepackage{tikz}
\newcommand*\circled[1]{\tikz[baseline=(char.base)]{
            \node[shape=circle,draw,inner sep=1pt] (char) {\small #1};}}

\begin{document}

\title{Loki: A System for Serving ML Inference Pipelines with Hardware and Accuracy Scaling}


\author{Sohaib Ahmad}
\affiliation{%
  \institution{University of Massachusetts Amherst}
  \country{USA}
  }
\email{sohaib@cs.umass.edu}

\author{Hui Guan}
\affiliation{%
  \institution{University of Massachusetts Amherst}
  \country{USA}
  }
\email{huiguan@cs.umass.edu}

\author{Ramesh K. Sitaraman}
\affiliation{%
  \institution{University of Massachusetts Amherst}
  \country{USA}
  }
\email{ramesh@cs.umass.edu}


\begin{abstract}
The rapid adoption of machine learning (ML) has underscored the importance of serving ML models with high throughput and resource efficiency. Traditional approaches to managing increasing query demands have predominantly focused on hardware scaling, which involves increasing server count or computing power. However, this strategy can often be impractical due to limitations in the available budget or compute resources. 
As an alternative, accuracy scaling offers a promising solution by adjusting the accuracy of ML models to accommodate fluctuating query demands. Yet, existing accuracy scaling techniques target independent ML models and tend to underperform while managing inference pipelines. Furthermore, they lack integration with hardware scaling, leading to potential resource inefficiencies during low-demand periods.
To address the limitations, this paper introduces \projectname{}, a system designed for  serving inference pipelines effectively with both hardware and accuracy scaling. \projectname{} incorporates an innovative theoretical framework for optimal resource allocation and an effective query routing algorithm, aimed at improving system accuracy and minimizing latency deadline violations. 
Our empirical evaluation demonstrates that through accuracy scaling, the effective capacity of a fixed-size cluster can be enhanced by more than $2.7\times$ compared to relying solely on hardware scaling. When compared with state-of-the-art inference-serving systems, \projectname{} achieves up to a $10\times$ reduction in Service Level Objective (SLO) violations, with minimal compromises on accuracy and while fulfilling throughput demands.

\end{abstract}

\maketitle

\section{Introduction}
\label{sec:intro}
The growing popularity of machine learning (ML) has led to the development of model serving systems\footnote{We use the terms ``model serving'' and ``inference serving'' interchangeably.}, where pre-trained ML models are hosted on a cluster of servers
to provide fast and accurate responses to inference queries. 
Model serving systems typically guarantee certain Service Level Objectives (SLOs) to users in terms of latency deadlines and, at the same time, strive to achieve high throughput and high resource efficiency in order to serve as many queries as possible in a given amount of time. 

As query demand (measured by queries per second or QPS) usually changes over time, model serving systems need to handle demand variations gracefully.
To accommodate increasing query demands, conventional methods primarily rely on hardware scaling, i.e., adding more devices or using more powerful accelerators, to improve system throughput~\cite{inferline, romero2019infaas}. 
However, hardware scaling may
not be feasible due to budget constraints or the limited availability of hardware resources in edge clusters or private clouds.

\textbf{Accuracy scaling.} Accuracy scaling has recently been proposed as an alternative to hardware scaling~\cite{ahmad2024proteus, guo2022sommelier}. 
A model serving system that uses accuracy scaling adapts model accuracy instead of hardware resources to gracefully handle query demand variations. 
Accuracy scaling is motivated by the fact that ML models can offer different levels of accuracy depending on the time spent computing the answer: less time spent on computation leads to less accurate results but higher throughput.
ML models with different accuracy profiles are called ``model variants'', which can be created from model compression techniques~\cite{sze2017efficient, blalock2020state}, or as part of neural network architecture designs.  
When the query demand is high, a model serving system with
accuracy scaling serves queries using less accurate model variants to avoid SLO violations. 
When the demand drops, the system serves queries using more accurate
model variants to improve accuracy.

Accuracy scaling is a particularly effective strategy in scenarios where query demands are so high that they risk overwhelming the available servers. In such cases, accuracy scaling ensures that the system continues to provide timely responses. These prompt responses that are less accurate are often more critical than slower more accurate responses, or worse, queries that fail to be processed at all. Accuracy scaling is especially beneficial for real-time ML applications, including interactive and cloud-based editing services, where quick feedback is essential~\cite{huang2015tencentrec, fang2018nestdnn}.

Accuracy scaling can be strategically combined with hardware scaling to optimize query handling. During periods of low query demand, the model serving system can employ hardware scaling to reduce the number of active servers, thereby optimizing resource usage. As query demand escalates, the system can shift to accuracy scaling. This transition enables the system to enhance its throughput capacity (in QPS), accommodating the surge in demand while still adhering to the SLOs set for user satisfaction.

\textbf{Limitations of existing approaches.}
The current approach to accuracy scaling is primarily tailored for serving individual, independent ML models. However, as ML becomes more integrated into practical applications, the use of \textit{inference pipelines} is increasingly common. These pipelines combine multiple ML models to tackle complex inference tasks and are becoming a standard part of ML inference workloads. For instance, an image generation pipeline such as Adobe Firefly \cite{adobefirefly} might sequentially employ a text embedding model, a diffusion model, and an image super-resolution model to produce high-resolution images from text prompts. 
The end-to-end inference latency of these pipelines must adhere to the specific SLOs set by the application (e.g., 200ms).  

When the existing accuracy scaling method~\cite{ahmad2024proteus} is applied to inference pipelines, it often leads to suboptimal resource allocation, resulting in high SLO violation rates and poor response quality. The core issue with the current accuracy scaling approach lies in its \textit{pipeline-agnostic} perspective on resource allocation. It adjusts the accuracy of ML models and allocates computing resources without considering the dependencies between the models in the different tasks of a pipeline. This lack of consideration for the inter-model relationships can impair the overall effectiveness of resource use in complex, multi-model inference tasks.

Another limitation of the existing accuracy scaling method is its lack of integration with hardware scaling. This shortcoming becomes evident particularly during periods of low query demand. In such scenarios, instead of scaling down the hardware resources, current methods continue to utilize all available servers to handle queries. This approach leads to inefficiencies, as it does not dynamically adjust the server usage based on the actual demand, resulting in unnecessary resource expenditure and under-utilization of the server infrastructure.

\textbf{The \projectname{} system.} 
To address the problem, this paper introduces \projectname{}\footnote{Loki is named after a Norse mythology figure who possessed the ability to change form and appearance, much like our system transitions between hardware and accuracy scaling.}, a model serving system designed to handle inference pipelines effectively using both hardware and accuracy scaling. 
The primary objectives of \projectname{} are to maximize the overall system accuracy and minimize the active server count, while adapting to fluctuating query demands. System accuracy is defined as the average accuracy across all queries processed by the system. 
\projectname{} operates under the assumption that, given sufficient resources, every query would prefer the most accurate model variant. However, it also recognizes that in situations where resources are constrained, a timely response with slightly lower accuracy is acceptable.

When query demand is relatively low compared to available server capacity, \projectname{} optimizes resource usage by reducing the number of active servers (and thus hardware costs) required to process queries with the most accurate model variants. Under these conditions, the system consistently achieves maximum accuracy. In contrast, as the query demand increases, \projectname{} smoothly transitions to a pipeline-aware accuracy scaling mode. This mode focuses on maximizing system accuracy while accommodating the increasing volume of queries, resulting in all servers actively participating in query processing.

\begin{figure}[t]
    \centering
        \centering
        \includegraphics[width=0.8\linewidth]{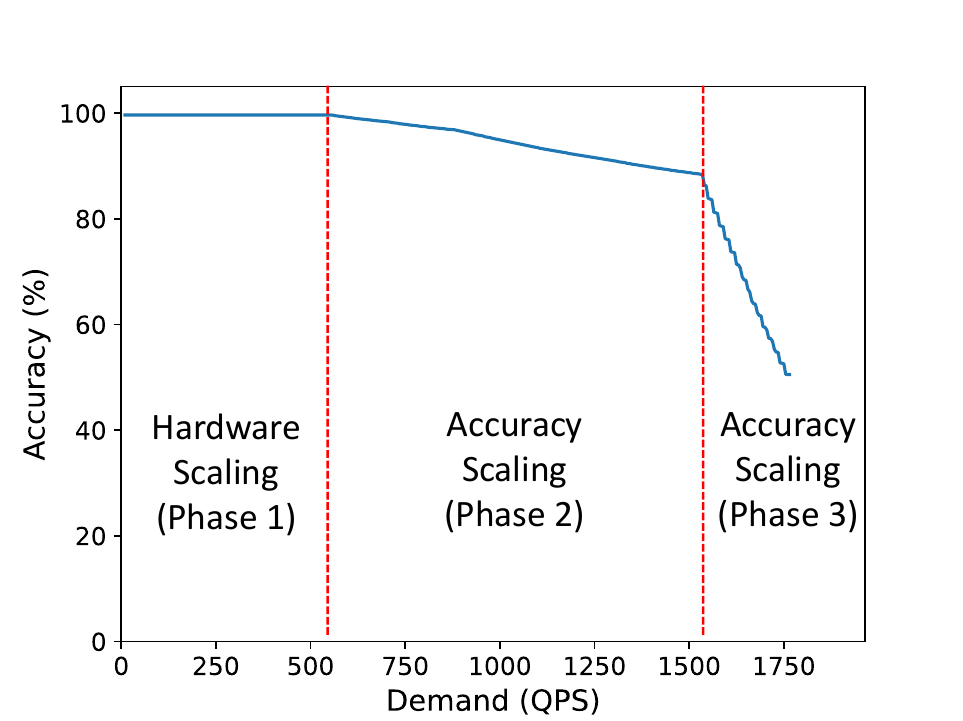}
        \caption{For a  traffic analysis pipeline that consists of two sequential tasks, Loki first accomodates the increase in query demand by using hardware scaling. 
        When demand increases further, Loki successively decreases the accuracy of each task of the pipeline to increase throughput to meet demand. Phase 2 decreases the $2^{\text{nd}}$ task's accuracy as it causes smaller end-to-end accuracy drop.}
    \label{fig:hw_vs_accuracy_scaling}
\end{figure}

{\bf Functioning of Loki.} To illustrate the functioning of Loki, we hosted a simple two-task traffic analysis ML pipeline on a  cluster of 20 servers. The first task  of the pipeline consists of an object detection model for identifying cars in an image and the second task classifies the identified car according to its make and model. Loki managed the resources in the server cluster using both hardware and accuracy scaling to serve user queries for this pipeline. 
 
 Figure~\ref{fig:hw_vs_accuracy_scaling} illustrates the functioning of Loki as it varied the system throughput to meet the user demand, ensuring that no queries were dropped.  
 In phase 1, as demand increased, Loki accommodated the demand by hardware scaling, while accuracy remained unchanged. 
 In this phase, Loki increased the number of available servers to meet the increasing query demand, but continued to use model variants that yield the highest accuracy.  
 In the second phase at around 560 QPS, it is no longer possible to scale the hardware since the server limits of the cluster have been reached. In response, Loki decreases accuracy to serve the increasing demand. Loki recognized that it is possible to get a larger increase in throughput for a given loss in end-to-end accuracy of the pipeline by using a less accurate model for the second task of the pipeline. 
 Consequently, it decreased the accuracy of the second task to increase throughput, while keeping the accuracy of the first task at the highest level. 
 As query demand continued to increase to about 1550 QPS, the system could no longer decrease the accuracy of task 2 to meet the demand. Therefore, it entered phase 3 where it starts to decrease the accuracy of task 1 of the pipeline. 
 This allows Loki to support upto 1765 QPS which is maximum throughput the system can support without dropping any request. 

It should be noted that Loki can support $2.7\times$ more throughput at the end of phase 2 than a system that does hardware scaling alone, albeit with a modest drop in accuracy of 13\%. Further, Loki can support up to $3\times$ more throughput at the end of phase 2 than hardware scaling alone, albeit with a more significant accuracy drop. In practice, there is usually a minimum level of acceptable accuracy required for queries, which limits the amount of accuracy scaling that can be performed.

{\bf Our contributions.} Our specific contributions follow: 
\begin{itemize}
    \item We design \projectname{}, the first model serving system that integrates hardware scaling with accuracy scaling to effectively serve inference pipelines. 
    \item We present a MILP-based theoretical framework for optimal allocation of resources in a cluster that incorporates performance models for the accuracy and throughput of inference pipelines. Using this framework, Loki periodically decides which model variants are hosted on which servers to meet throughput, accuracy, and latency requirements.
    \item When queries arrive, they need to be routed to the right sequence of model variants in the pipeline. For this task, we propose an efficient routing algorithm that intelligently routes the queries in real-time to maximize system accuracy and minimize SLO violations.
    \item We evaluate \projectname{} against two state-of-the-art model serving systems, one system that performs hardware scaling but not accuracy scaling~\cite{inferline} and another that performs accuracy scaling but is pipeline-agnostic~\cite{ahmad2024proteus}. Using query workloads from synthetic and production traces, we show that \projectname{} reduces SLO violations by more than $10\times$ compared to pipeline-agnostic accuracy scaling systems while using $2.7\times$ fewer servers during off-peak times where the demand is low. Further, Loki increases the effective capacity of the cluster by more than $2.7\times$ compared to systems that do not perform accuracy scaling.
\end{itemize}


\section{Background and Challenges}
\label{sec:motivation}
This work is motivated by the importance of serving inference pipelines and the benefits of accuracy scaling in model serving. We provide the background and then outline the challenges in building a pipeline-aware inference serving system.

\subsection{Background}
\label{sec:background}

\textbf{Inference pipelines.} 
Inference pipelines integrate multiple ML tasks together in a dataflow graph to address more complex tasks.
These pipelines can be represented as \textit{directed rooted trees}, where each node (or vertex) represents a task, the input, or the output, and each directed edge of two tasks denotes the flow of data between them\footnote{Loki does not support general directed acyclic graphs where an ML model derives input from multiple models. This paper uses the terms ``inference pipeline'' and ``pipeline graph'' interchangeably.}.
The root of the tree is referred to as the source (i.e., the input) and the leaves of the trees are the sinks (i.e., the outputs). 
Thus, the  rooted tree consists of multiple source-to-sink paths, where each of these paths has its own end-to-end accuracy. The end-to-end accuracy of the pipeline graph is the average of the end-to-end accuracy of all the source-to-sink paths.

\begin{figure}[t]
    \centering
    \begin{subfigure}{0.52\linewidth}
        \centering
        \includegraphics[width=\textwidth]{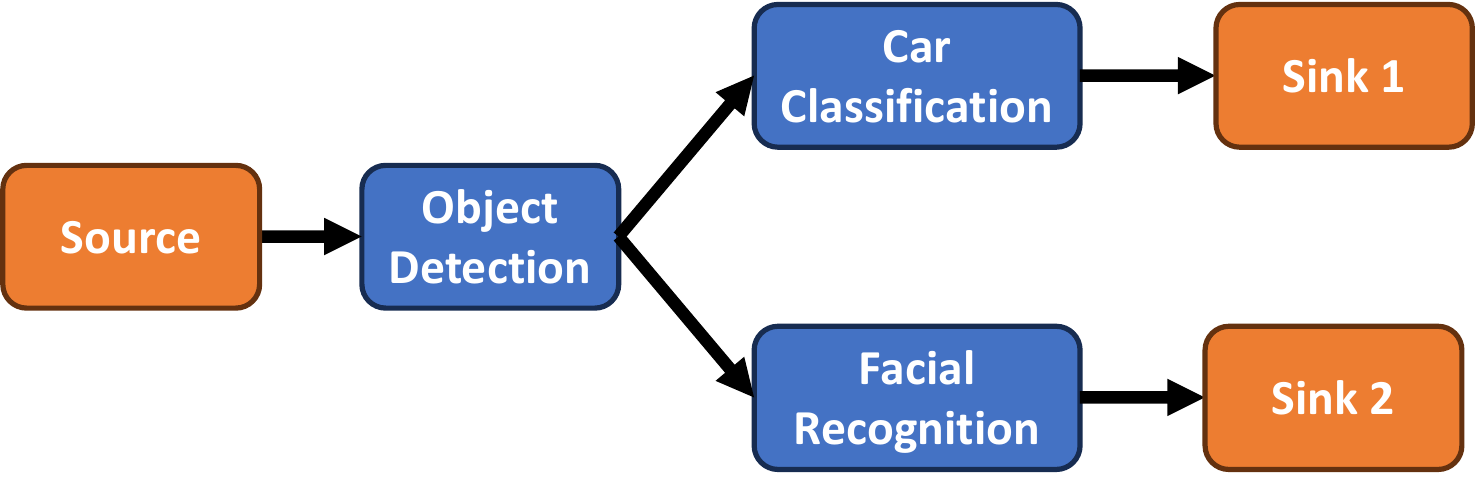}
        \caption{Traffic analysis pipeline}
    \end{subfigure}
    \begin{subfigure}{0.39\linewidth}
        \centering
        \includegraphics[width=\textwidth]{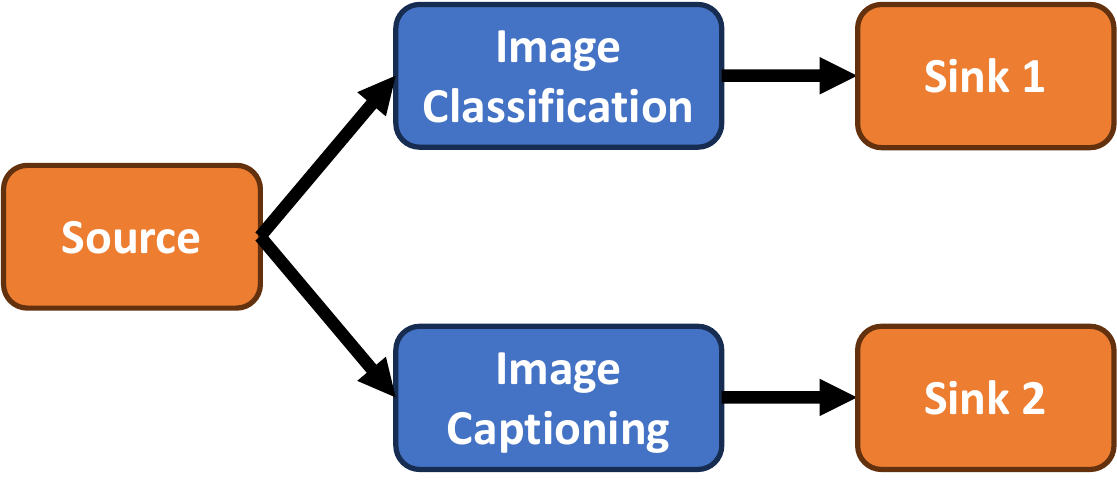}
        \caption{Social media pipeline}
    \end{subfigure}
    \caption{Examples of inference pipelines
    }
    \label{fig:inference-pipelines}
\end{figure}

In the execution of an inference pipeline for serving a query (also called a request\footnote{We use the terms query and request interchangeably in this work.}), the ML model for one task generates intermediate outputs that serve as inputs (termed \textit{intermediate queries}) for the ML model in the subsequent tasks. Figure~\ref{fig:inference-pipelines} illustrates two representative inference pipelines studied in this paper.
The traffic analysis pipeline can be used to generate traffic analytics on the video feed from cameras at intersections. 
The social media pipeline can be used by platforms such as Twitter and Facebook to detect objects in the image and generate suggested captions.

\begin{figure}[t]
    \centering
        \centering
        \includegraphics[width=0.6\linewidth]{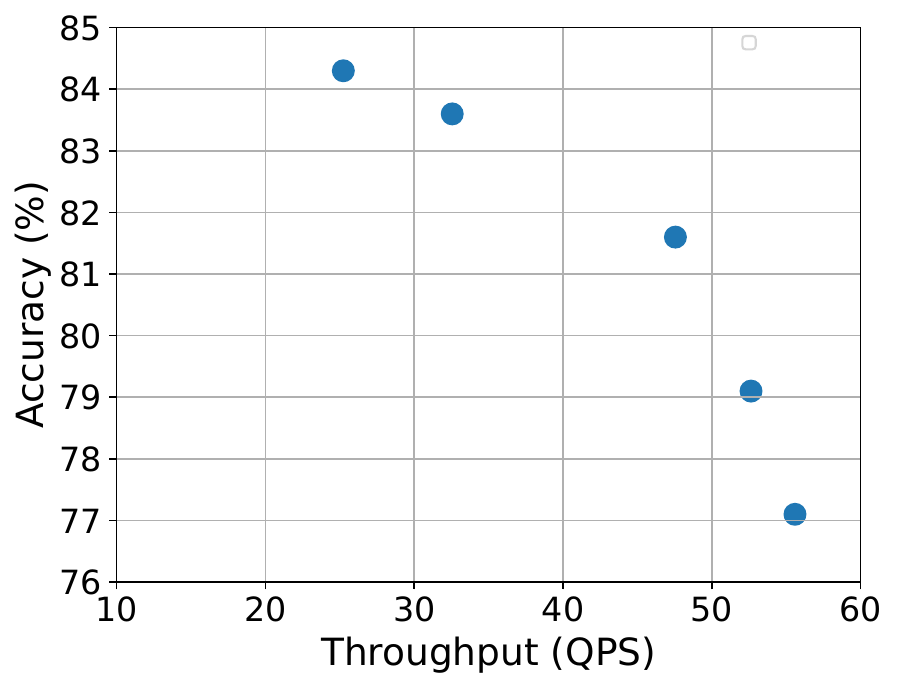}
        \caption{Accuracy-throughput tradeoff for EfficientNet model variants as profiled on an NVIDIA V100 GPU}
    \label{fig:accuracy_throughput_tradeoff}
\end{figure}

\textbf{Accuracy scaling.}
Accuracy scaling leverages the fact that model variants with different compute complexities (e.g., models from the EfficientNet family~\cite{tan2020efficientnet}) can be used to serve the same inference task.
A model variant that is more lightweight is usually less accurate, but can be executed faster, resulting in higher throughput on the same hardware, as shown in Figure~\ref{fig:accuracy_throughput_tradeoff}. 
The concept of accuracy scaling was first introduced in Proteus~\cite{ahmad2024proteus}, a model serving system designed for handling independent ML models on a cluster with a fixed number of servers. Accuracy scaling is particularly effective in managing high query demands with a limited number of servers. In scenarios where the volume of queries exceeds the server capacity, accuracy scaling strategically reduces the accuracy of the models. This reduction is done to ensure that the system meets the latency deadlines of the queries, thus balancing the trade-off between accuracy and timely response under heavy load conditions.

\subsection{Challenges}
Despite the promise of accuracy scaling, applying it to serve inference pipelines is challenging due to the complexities introduced by the inter-dependencies of ML models.

\subsubsection{Optimal resource allocation.}
\label{sec:challenge_allocation}
In this work, a resource allocation plan includes three key specifications: (1) the choice of model variants for each task of an inference pipeline, 
(2) the number of replicas for each model variant (termed \textit{replication factor}), 
and (3) the maximum batch size that can be used for each model variant. 
The maximum batch size corresponds to the maximum time budget assigned to a task. 

In the context of accuracy scaling, an optimal resource allocation plan maximizes system accuracy while satisfying a target query demand given a fixed cluster size. The accuracy scaling approach introduced in Proteus~\cite{ahmad2024proteus} is pipeline-agnostic, meaning it adjusts the accuracy of ML models individually without considering the interdependencies between them. When applied to inference pipelines, this approach can lead to suboptimal resource allocation, resulting in poor query response quality and high rates of SLO violations. These interdependencies present three major issues:

\textit{1. Impact of the accuracy of individual models on the end-to-end accuracy of the pipeline.}
   Choosing model variants for each task must be made with the knowledge of its impact on the end-to-end accuracy on the pipeline.  When facing increased query demands, the system should reduce the accuracy of models that minimally affect the end-to-end pipeline accuracy. For instance, Figure~\ref{fig:hw_vs_accuracy_scaling} shows that decreasing accuracy of the second task in the pipeline causes smaller end-to-end accuracy drop compared to the first task. 
   This consideration is absent in the existing accuracy scaling methods, which do not consider the influence of individual models on end-to-end pipeline accuracy. 

\textit{2. Throughput bottlenecks.}
   The optimal batch size and replication factor for each selected model variant depend on the throughput bottleneck of the inference pipeline. If a given task is not the bottleneck, increasing the batch size or assigning more resources to a model variant of that task enhances throughput for that task but does not necessarily improve overall system throughput. Moreover, allocating more resources to the bottleneck task may create resource shortages for other tasks, potentially creating new bottlenecks. Using a larger batch size at a given task also introduces longer processing delays for that task, reducing the time available for other tasks. This is a departure from the scenario addressed by Proteus, where ML models are independent and throughput improvements in any model enhance overall system throughput.

\textit{3. Workload multiplication effects.}
   The workload of each task can be influenced by the model variant used in preceding task. For example, in a pipeline with a face detection model followed by a face recognition model, the input demand for the recognition model depends on the output of the detection model. A more accurate detection model might detect more faces, thereby increasing the workload for the recognition task. The existing accuracy scaling approach fails to account for such workload dependencies between models in resource management. 

\textbf{Our approach.} 
We design performance models that assess how a resource allocation plan influences system accuracy, latency, and throughput capacity. 
These performance models are particularly crafted to consider the intricate relationships between the models in a inference pipeline. Utilizing these performance models, we can frame the resource allocation problem within a Mixed-Integer Linear Programming (MILP) framework and leverage MILP solvers to determine the optimal resource allocation plan.
\projectname{} periodically invokes the solver to re-allocate resources to accomondate macro-scale query demand changes.

Additionally, the performance models enable us to integrate hardware scaling into the same MILP framework used for accuracy scaling, albeit with a distinct optimization goal. In terms of hardware scaling, the ideal resource allocation plan is defined as the one that minimizes the number of active servers required to process queries while meeting a target query demand. The optimization formula for this objective is based on our performance model, which delineates the connection between the system's latency, throughput capacity, and the specifics of a resource allocation plan. This unified approach under the MILP framework allows for a cohesive treatment of both hardware and accuracy scaling, each with its unique optimization targets, while maintaining a consistent underlying methodology. We present the performance models and our optimization in Section~\ref{sec:resource_manager}.

\subsubsection{Query execution with accuracy scaling.}
\label{sec:challenge_routing}
Another challenge in utilizing accuracy scaling for serving inference pipelines is deciding the optimal execution path for each incoming query. This decision aims to enhance system accuracy while minimizing violations of SLO. The MILP formulation used for system accuracy estimation operates under the assumption that each deployed ML model functions at its maximum throughput to meet the target query demand. However, this assumption may not always be valid due to the dynamic nature of query arrivals during runtime. The method by which queries are routed and executed can significantly affect both the quality of their responses and their capacity to adhere to predetermined latency deadlines.

\textbf{Our approach.}
We present a greedy request routing algorithm in Section~\ref{sec:load_balancer} that routes requests in a way that maximizes system accuracy. To minimize SLO violations, we perform a runtime optimization to drop requests that are unlikely to meet their SLOs, in order to free up resources for requests with higher chances of meeting their SLOs.


\section{System Architecture of Loki}
\label{sec:architecture}
\begin{figure}[t]
    \centering
        \centering
        \includegraphics[width=\linewidth]{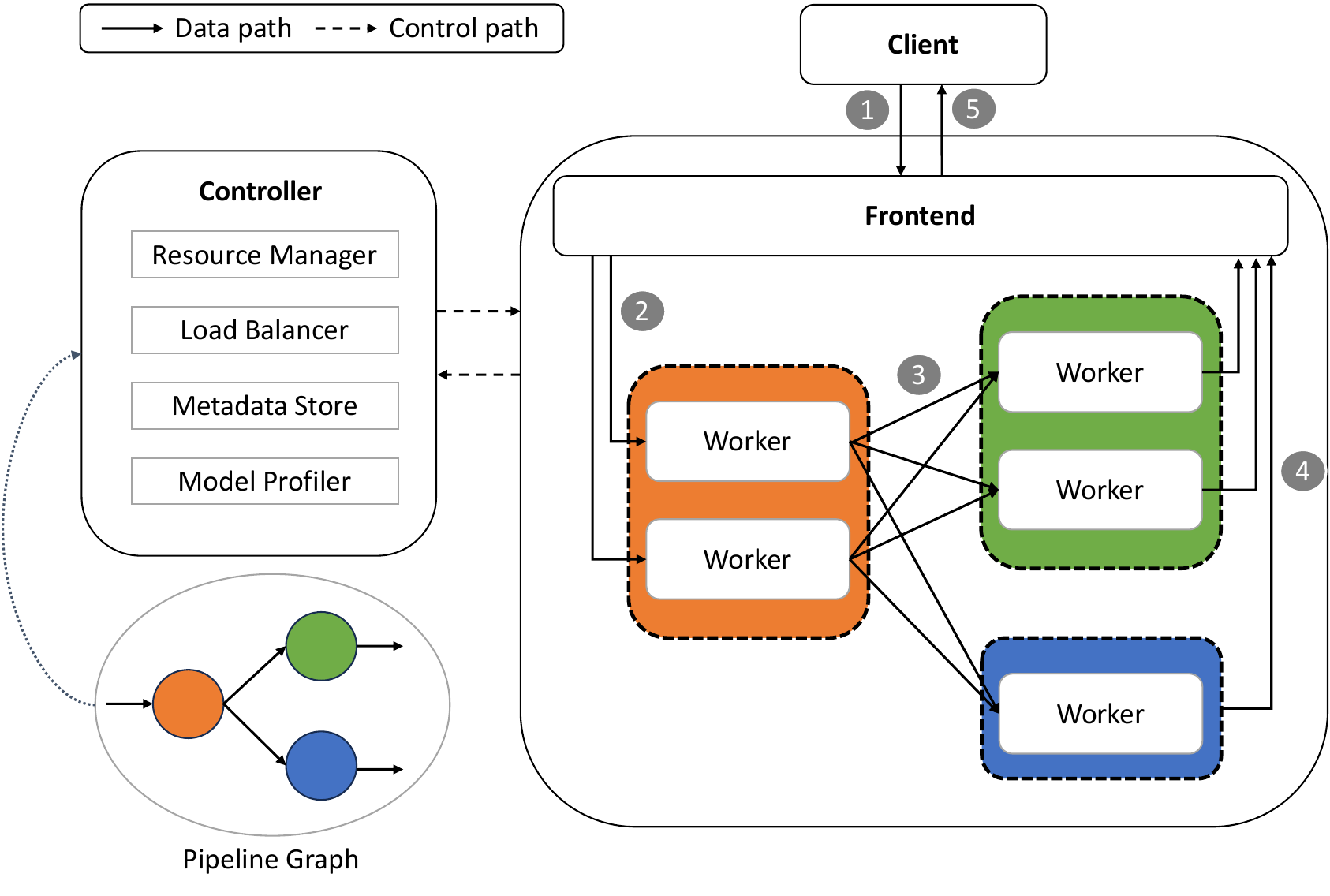}
        \caption{System Architecture of Loki}
    \label{fig:system_architecture}
\end{figure}

We now present an overview of Loki's system architecture and provide more details in Sections~\ref{sec:resource_manager} and ~\ref{sec:load_balancer}. Figure~\ref{fig:system_architecture} shows the three key components: Controller, Frontend, and Workers.

\textbf{Controller.} The Controller is responsible for managing the resources in the system and for routing the queries. It uses the following sub-components to achieve this.

\textit{Resource Manager.} The Resource Manager performs resource allocation periodically in response to the incoming demand to indicate which model variants to host, as well as their replication factors and maximum batch sizes. It consults the Metadata Store to get the historical query demand, the pipeline graph, and the profile of model variants for each task in the graph to perform the allocation. Once the Resource Manager obtains an allocation plan, it adjusts the allocation of workers to model variant instances in the system to reflect the new allocation plan.  
The Resource Manager assumes a finite-size cluster for allocation. As long as it can meet demand using the highest accuracy model variants for each task in the pipeline, it tries to scale the hardware needed to serve the demand. If the demand cannot be met even using the entire cluster, it drops accuracy to meet the increased demand. We explain the details of the resource allocation algorithm in Section~\ref{sec:resource_manager}.

\textit{Load Balancer.} The Load Balancer is tasked with routing the queries through the hosted instances to maximize system accuracy. It uses the resource allocation plan set up by the Resource Manager as well as the pipeline graph and real-time demand from the Metadata Store to perform the routing. It sets up routes from the Frontend to the first-task workers of the pipeline, as well as the routes between intermediate workers.

\textit{Model Profiler.} The Controller uses the Model Profiler to profile the expected processing times of each model variant in the pipeline with different batch sizes during the initial setup. The profiles are then stored in the Metadata Store and used by the Resource Manager every time it performs resource allocation.

\textit{Metadata Store.} The Metadata Store holds the information required by the Resource Manager and Load Balancer. It stores the representation of the pipeline as a graph, the profiled throughput and accuracy of each model variant, and the profiled end-to-end accuracy of each source-to-sink path through the graph.
During the initial setup, a pipeline graph, the model variants for each node in the graph, and the end-to-end pipeline latency requirement are registered in the Metadata Store.

\textbf{Frontend.} The Frontend accepts queries from the client and forwards them to the respective workers. The query is then forwarded by those workers to intermediate workers in the pipeline, and the workers at the last task of the pipeline return the results to the Frontend which then returns the results to the client. The Frontend also records the incoming demand into the system and reports it to the Controller which stores it in the Metadata Store.

\textbf{Workers.} The workers host the model variants and execute inference queries. Each worker has a queue that it uses to form batches. As the worker executes queries, it records the number of subsequent requests generated for downstream tasks in terms of a multiplicative factor on the incoming number of requests and reports it to the Controller using heartbeat messages.

\textbf{Query Processing.} Clients interact with \projectname{} in the following way. Client sends a query to the Frontend of \projectname{} (\circled{1}). The Frontend routes the query to one of first-task workers (\circled{2}). The first-task worker passes the intermediate query (or queries) to one (or more) second-task worker (\circled{3}) and so on. The last-task worker (or workers) pass the inference results to the Frontend (\circled{4}). The Frontend aggregates the results and returns them to the client (\circled{5}). 

We now provide details of the two core modules: the Resource Manager and the Load Balancer in Sections~\ref{sec:resource_manager} and \ref{sec:load_balancer} respectively.


\section{Resource Manager}
\label{sec:resource_manager}
The Resource Manager is tasked with allocating resources in the system to meet the incoming demand. It uses the incoming demand as input and outputs the resource allocation plan that describes the model variants to host as well as the replication factor and the maximum batch size that can be used for each model variant by performing the following two steps.

\begin{enumerate}
    \item \textbf{Hardware scaling.} The Resource Manager first tries to serve the incoming demand with the minimum number of workers by using the most accurate model variants for each task in the pipeline. If this is not possible, it executes the accuracy scaling step below.
    \item \textbf{Accuracy scaling.} If the Resource Manager is unable to meet demand by using the entire cluster with the most accurate model variants, it tries to determine the minimum amount of system accuracy to sacrifice in order to meet the demand. This enables the Resource Manager to increase the throughput capacity of the cluster, allowing it to serve a greater demand compared to using hardware scaling alone.
\end{enumerate}

Each of the above steps is modeled as a mixed-integer linear program (MILP) as described below. The MILPs are solved by the Resource Manager to get the resource allocation plan.

\subsection{MILPs for hardware and accuracy scaling}
We now formulate the resource allocation problem as a mixed integer linear programming (MILP) optimization. 
We first elaborate the input and the output of the optimization problem and then introduce the performance models that quantify the relationship between a resource allocation plan and system accuracy, latency, and throughput. 
Table~\ref{table:notation} summarizes the notation used in the optimization.

\begin{table}[t]
    \tabcolsep=0.05cm 
    \small 
    \begin{center}
    \begin{tabular}{ c p{6.5cm}}
     \toprule 
     \multicolumn{2}{l}{\underline{Subscripts}} \\
     $T$ & the set of tasks \\
     $t_i$ & the $i^{\text{th}}$ task in the pipeline, $t_i \in T$ \\
     $V_{i}$ & the set of model variants for the $i^{\text{th}}$ task \\
     $v_{i,k}$ & the $k^{\text{th}}$ model variant for the $i^{\text{th}}$ task, $v_{i,k} \in V_i$ \\
     $E$ & the set of edges between tasks in the pipeline graph \\
     $P$ & the set of all root-to-sink paths in the augmented graph \\
     $B$ & the set of allowed batch sizes \\
     $b$ & batch size, $b \in B$ \\
     
      \multicolumn{2}{l}{\underline{Inputs}} \\
      $D$ & incoming demand (QPS) \\
      $S$ & number of workers in the cluster \\
      $L$ & latency SLO of the pipeline \\
      $r(i,k)$ & multiplicative factor for the $k^{\text{th}}$ model variant of $i^{\text{th}}$ task\\
      $q(i, k, b)$ & profiled throughput (QPS) for the $k^{\text{th}}$ model variant of the $i^{\text{th}}$ task with batch size $b$ \\
      $A(v_{i,k})$ & profiled accuracy of the $k^{\text{th}}$ model variant of the $i^{\text{th}}$ task \\
      $\hat{A}(p)$ & end-to-end profiled accuracy of path $p$ \\
    \multicolumn{2}{l}{\underline{Optimization variables}} \\ 
    $x(i,k)$ & number of instances for the $k^{\text{th}}$ model variant of the $i^{\text{th}}$ task \\
    $y(i, k)$ & maximum batch size to use for the $k^{\text{th}}$ model variant of the $i^{\text{th}}$ task \\
    
     \multicolumn{2}{l}{\underline{Intermediate variables}} \\
     $c(p)$ & ratio of queries supported through path $p$ \\
     $I(p)$ & 1 if there is any traffic through the path $p$; 0 otherwise \\
     \multirow{2}{*}{$l(i,k)$} & the processing latency of the $k^{\text{th}}$ model variant of the $i^{\text{th}}$ task with the configured batch size \\
     $\hat{l}(p)$ & end-to-end latency through path $p$ \\
     \bottomrule  
    \end{tabular}
    \end{center}
    \caption{Notation for MILP}
    \label{table:notation}
    \vspace{-0.8cm}
\end{table}

\textbf{Inputs.} We are given as input the pipeline graph consisting of a set of tasks $T$ and a set of directed edges $E$ where an edge $e = (i, j) \in E$ denotes an edge from the $i^{\text{th}}$ task $t_i$ to the $j^{\text{th}}$ task $t_j$. The pipeline graph is a directed rooted tree with the source node ($t_1$) as the root which does not have any incoming edges. We are also given the incoming demand of the system, $D$, that arrives at the root. Let $r(i,k)$ represent the multiplicative factor of the $k^{\text{th}}$ model variant of task $t_i \in T$.

\textbf{Output.} The output is the resource allocation plan, defined by the two optimization variables: $x(i, k)$ and $y(i, k)$, representing the number of instances to host for the $k^{\text{th}}$ model variant of task $t_i \in T$ along with the maximum batch size to use for it, respectively.

\textbf{Meeting the system throughput demand.} To model how an allocation plan affects the system throughput, we need to introduce two concepts: augmented graph, and intermediate query demand. 

\textit{Augmented graph.} The augmented graph aims to represent all possible materializations of a pipeline graph using different combinations of model variants for each task. We construct an augmented graph from the given pipeline graph in the following way: For every vertex $i$ in the pipeline graph that represents the $i^{\text{th}}$ task $t_i$, we create vertices $(i,k)$ in the augmented graph representing the $k^{\text{th}}$ model variant of task $t_i$. We add a directed edge from a vertex $(i,k)$ to $(j,k')$ in the augmented graph if $(i,j)$ is an edge in the pipeline graph, for all $k,k'$.

\textit{Intermediate query demand.} The Resource Manager not only needs to host enough model instances to serve the incoming queries at the first task of the pipeline, but it also needs to consider the intermediate queries generated by the initial tasks to host model instances for the downstream tasks in the pipeline. For example, an object detection model in the traffic analysis pipeline may detect 10 cars in an image, creating 10 intermediate queries to be served by the car classification model. Therefore, it needs estimates of the multiplicative factor for each model variant. The Resource Manager uses the estimate of incoming demand into the system as well as the profiled multiplicative factor of each model variant to estimate the intermediate query demand.

We next model the requirement that model variants chosen for a task need to meet the task's intermediate query demand in  Constraint~\ref{eq:path-constraint-1}.
For every model variant $v_{i,k}$, we want to ensure that it has enough resources to serve all requests arriving at it.
To get the number of requests arriving at $v_{i,k}$, we need to consider all paths that contain it.
Let $P'_{i,k}$ be the set of all paths $p$ that start at a vertex that corresponds to the root and end in vertex $(i,k)$ that represents model variant $v_{i,k}$. 
For $p \in P'_{i,k}$, let $m(p, i, k)$ represent the number of requests derived from a single request entering path $p$ that reach $v_{i,k}$. Thus, the following holds.
\begin{align}
    m(p, i, k) &= \prod_{(i',k') \in p} r(i',k') \label{eq:partial_path_mult_factor}
\end{align}

We now ensure that $v_{i,k}$ has enough resources to serve all requests going through it. Let $P$ be the set of all paths in the augmented graph that start at a vertex that corresponds to the root and end at a vertex that corresponds to a sink. Let $P_{i,k}$ be the set of all paths $p \in P$ that include vertex $(i,k)$ that represents model variant $v_{i,k}$. 
As the total number of requests per second entering the pipeline is $D$ and $c(p)$ is the ratio of these requests that we route through the path $p$, the number of requests that arrive at $v_{i,k}$ after multiplication are $\sum_{p \in P_{i,k}} D \cdot c(p) \cdot m(p, i, k)$. Then, to ensure $v_{i,k}$ has enough resources to serve all the requests going through it, we add the following constraint.

\begin{align} 
    \sum_{p \in P_{i,k}} D \cdot c(p) \cdot m(p, i, k) &\leq x(i,k) \cdot q(i, k, y(i,k))  &\forall v_{i,k} \in V_i, \forall t_i \in T  \label{eq:path-constraint-1}
\end{align}

We require that the number of workers used may not exceed the total available workers in the cluster.

\begin{align}
    \sum_{i,k} x(i,k) &\leq S   \label{eq:hw-constraint-1}
\end{align}

\textbf{Meeting the latency SLO.}
We now model how a resource allocation plan affects the end-to-end pipeline latency, which is bounded by the SLO requirements of queries.  
The maximum batch size is bounded by the latency requirements of requests and also the largest batch size feasible on a specific device. 
We configure the maximum batch size using one of the batch sizes from the set of allowed batch sizes (Constraint~\ref{eq:batch-size-constraint-1}). The processing latency of a model variant depends on the maximum batch size configured for it and the throughput of the model using that batch size (Constraint~\ref{eq:latency-constraint-1}).

\begin{align} 
    y(i,k) &\in B  &\forall v_{i,k} \in V_i, \forall t_i \in T \label{eq:batch-size-constraint-1} \\
    l(i,k) &= \frac{y(i,k)}{q(i,k, y(i,k))}  &\forall v_{i,k} \in V_i, \forall t_i \in T  \label{eq:latency-constraint-1}
\end{align}

We define the end-to-end processing latency through a path $p$ as following.

\begin{align}
    \hat{l}(p) &= \sum_{(i,k) \in p} l(i,k) &\forall p \in P
\end{align}

As we need to meet the end-to-end latency SLO of the pipeline, we need to ensure that the processing latency through each path serving any query is less than the SLO.

\begin{align}
    \hat{l}(p) \cdot I(p) &\leq L &\forall p \in P \label{eq:latency-constraint-2}
\end{align}

To account for the waiting time of queries in the queue, we divide the SLO by two. This is motivated by an observation from prior work~\cite{shen2019nexus, ahmad2024proteus}: a query that arrives right after a batch starts executing needs to wait for the current batch to finish, before starting execution with the next batch and thus may have to wait twice the amount of processing time of a batch.
Based on this, we divide the SLO by two, assuming that the query's waiting time in the queue is as long as the query's execution time.

\textbf{Modeling the system accuracy.} 
As mentioned in Section~\ref{sec:challenge_allocation}, the model variants chosen at each task of the pipeline affect the end-to-end accuracy. Therefore, to capture this accuracy, we profile the end-to-end accuracy of every path $p \in P$ as $\hat{A}(p)$. Given that the optimization configures the ratio of requests through the path $p$ to be $c(p)$, the system accuracy is $\sum_{p \in P} c(p) \cdot \hat{A}(p)$.

\textbf{The MILP optimization.} 
We now present the MILP optimization for both hardware and accuracy scaling based on the above-mentioned performance models. 

\textit{Step 1: Hardware scaling.}
We first try to serve demand using the most accurate model variant for each task. To achieve this, we constrain the number of hosted instances for all other model variants to be 0. Let us denote the most accurate model variant for the task $t_i$ as:

\begin{align}
    v_{i}^{max} &= \underset{v_{i,k} \in V_i}{\arg\max} A(v_{i,k}) &\forall t_i \in T 
\end{align}

Then we can denote the set of all other model variants as:

\begin{align}
    \overline{V_i} &= \{ v_{i,k} \in V_i | A(v_{i,k}) < A(v_{i}^{max}) \} &\forall t_i \in T
\end{align}

We can now define the constraint to disallow less accurate model variants as following.

\begin{align}
    x(i,k) &= 0   &\forall v_{i,k} \in \overline{V_i}, \forall t_i \in T \label{eq:highest_acc_only}
\end{align}

In this case, the optimization objective is to minimize the number of workers used to serve the demand.

\begin{align}
    min \sum_{i,k} x(i,k) \quad \text{ s.t. Constraints \ref{eq:partial_path_mult_factor}-\ref{eq:highest_acc_only} hold}  \label{eq:hw-scaling-objective}
\end{align}

It is important to note that it may not be possible to serve demand with the highest accuracy model variants by using even the entire cluster. In this case, the above optimization will immediately detect the constraints to be infeasible, and we resort to accuracy scaling.

\textit{Step 2: Accuracy scaling.}
The optimization objective for accuracy scaling is to maximize the system accuracy, which is the average accuracy experienced by all queries served by the system. The system accuracy is measured by multiplying the end-to-end accuracy of each path by the ratio of queries flowing through it.

\begin{align}
    max \sum_{p \in P} c(p) \cdot \hat{A}(p) \quad \text{ s.t. Constraints \ref{eq:partial_path_mult_factor}-\ref{eq:latency-constraint-2} hold}      \label{eq:acc-scaling-objective}
\end{align}

\subsection{Solving the MILP}
As the Resource Manager is invoked periodically to respond to long-term changes in demand, it can tolerate a higher runtime from considering a large number of paths through the pipeline, as long as it yields an optimal solution at the end. For the purpose of our experiments, we use a 10-second invocation interval for the Resource Manager. We show in Section~\ref{sec:milp_overhead} that the runtime overhead of the MILP is low enough to allow it to adapt reasonably quickly to this invocation frequency. Additionally, the Resource Manager may reallocate resources if it detects a significant change in the demand between its periodic invocations. To estimate the demand to serve, we use an exponentially weighted moving average on the recent demand history.

\textbf{Latency budgets for tasks.} The batch sizes set by the MILP not only serve as guidance for the workers to form batches during execution, they also allow us to set latency budgets for each task. Since the optimization ensures that the execution latency through every path falls under the SLO using the configured batch sizes, we use the execution time of a model variant with the configured batch size as the latency budget for its task. These latency budgets are useful during query execution to make sure requests are on track to meet their SLOs as they move through the tasks in the pipeline. In case they fall behind, we can use the latency budgets to perform early dropping of requests as we detail in Section~\ref{sec:early_dropping}.

\textbf{Communication latency.} It is important to consider the communication latency between workers since the end-to-end execution latency of a query depends on the communication latency between the workers that serve that particular query. As we consider all servers to be in the same cluster, we assume communication latency between any pair of servers to be homogeneous. Therefore, during resource allocation, we subtract the product of the total number of servers in the path with this communication latency from the latency SLO of the pipeline.

\textbf{Estimating multiplicative factors.} As mentioned before, each request generates multiple requests for downstream tasks in the pipeline. We refer to the number of outgoing requests generated for each incoming request as the multiplicative factor. We note that every model variant can have a different multiplicative factor, for example, extending the example from above, a lower accuracy object detection model such as YOLOv5n may detect fewer cars in an image compared to a higher accuracy model variant such as YOLOv5x.
Each model variant hosted at a worker records the multiplicative factors it observes when serving queries and reports them to the Controller through heartbeat messages. The Controller aggregates these for each model variant to be used by the Resource Manager.


\section{Load Balancer}
\label{sec:load_balancer}

The Load Balancer produces the routing tables that enable each query to be routed through a sequence of model variants in real-time to maximize system accuracy and reduce SLO violations. To achieve this, it takes as input the resource allocation plan produced by the Resource Manager, the pipeline graph, and the recent demand history from the Metadata Store, and outputs routing tables for both the Frontend and workers. 

The Load Balancer is a centralized component and periodically updates the routing tables of workers, while the workers use their respective routing tables during real-time execution to find downstream workers for intermediate requests.
We present our algorithm in this section and explore the overhead of the Load Balancer in Section~\ref{sec:milp_overhead}.

\subsection{Request Routing}
\label{sec:request_routing}

We now present our request routing algorithm, \textsc{MostAccurateFirst} (Algorithm~\ref{alg:online_heuristic}).
The algorithm works in the following way: Starting from the root node of the pipeline graph, it takes the incoming QPS of the node and assigns model variants to it in non-increasing order of their profiled single-model accuracies. As each model variant can have a different multiplicative factor, the outgoing requests for this node are calculated by multiplying the requests assigned to each model variant by the multiplicative factor of that variant. The outgoing requests are sent to the children nodes, and we recursively repeat the same procedure on each of the children. \textsc{MostAccurateFirst} generates routing tables based on estimated demand and updates the routing tables of all the workers and the Frontend.

\renewcommand{\algorithmicindent}{0.5em}
\begin{algorithm}
\caption{}
\label{alg:online_heuristic}
\begin{algorithmic}[1]

\Procedure{MostAccurateFirst}{pipelineGraph, worker metadata}
\State sortedGraph $\gets$ \textsc{TopologicalSort}(pipelineGraph)
\State routingTables $\gets \phi$
\For{task in sortedGraph}
    \State workers $\gets$ sort(task.workers) \Comment{By single-model accuracy}
    \For{worker in workers}
        \State workerTable $\gets \phi$
        \For{child in task.children}
            \State outgoing $\gets$ worker.incoming * task.multFactor * child.branchRatio
            \State totalChildDemand $\gets$ outgoing
            \State childWorkers $\gets$ sort(child.workers)
            \For{cWorker in childWorkers}
                \If{cWorker.capacity > 0 \& outgoing > 0}
                    \State routed $\gets$ min(outgoing, cWorker.capacity)
                    \State routingProbability $\gets$ routed / totalChildDemand
                    \State workerTable.addEntry(cWorker, routingProbability)
                    \State outgoing $\gets$ outgoing $-$ routed
                    
                    \State cWorker.capacity $\gets$ cWorker.capacity $-$ routed
                    \State cWorker.incoming $\gets$ cWorker.incoming + routed
                    
                \EndIf
            \EndFor
        \EndFor
        \State routingTables[worker] $\gets$ workerTable
    \EndFor
\EndFor
\State return routingTables
\EndProcedure
\end{algorithmic}
\end{algorithm}

The Load Balancer runs the \textsc{MostAccurateFirst} algorithm every time the Resource Manager changes the resource allocation plan. It also runs periodically between successive invocations of the Resource Manager. On each execution, the \textsc{MostAccurateFirst} algorithm produces routing tables for every worker and pushes the routing tables to the respective workers. The workers then use their routing tables during real-time query execution to find downstream workers to forward intermediate queries. Since we saturate workers for each node in non-increasing order of their single-model accuracies, we may have some workers for each node with leftover capacity. We make a list of these workers along with their leftover capacities and propagate this list to their upstream workers. The upstream workers can use this list to perform Opportunistic Rerouting, a technique we describe in Section \ref{subsec:opp_rerouting}.

As the end-to-end pipeline accuracy is a monotonic function of single-model accuracies, and \textsc{MostAccurateFirst} ensures that each node in every source-to-sink path in the pipeline graph gets the highest single-model accuracy for a given QPS, this means \textsc{MostAccurateFirst} maximizes the end-to-end pipeline accuracy.

\subsection{Early dropping with opportunistic rerouting}
\label{sec:early_dropping}

The Resource Manager and Load Balancer assess the recent demand history of the system to allocate resources and set up routing for future requests. However, real-time demand can deviate from these estimates. Moreover, the estimates are made at the granularity of seconds, while request arrivals and multiplicative factors may fluctuate at smaller timescales between these estimates. Due to these reasons, there is a possibility that some requests may exceed their SLOs despite provisioning the system to prevent SLO violations. In such instances, it may be more effective to preemptively drop a request if we can anticipate that it is likely to miss its SLO. This decision can only be made at runtime during query execution at individual workers. We refer to this process as \textit{early dropping}, and it can mitigate SLO violations by freeing up resources for requests that are expected to meet their SLOs.

Since requests undergo execution sequentially through the tasks within the pipeline, we use the latency budget of each task to estimate whether a request is on track to meet its SLO. Recall that we set the latency budget of each task by using the batch sizes set by the Resource Manager for each hosted model variant.

We consider two naïve mechanisms to perform early dropping using the allocated latency budgets for the pipeline tasks:

\begin{enumerate}
    \item \textbf{Per-task dropping.}
    When a request finishes execution at a given task, we note the total time spent by the request at the task, i.e., the processing time of the request as well as the time it spent waiting in the queue. If this time exceeds the latency budget assigned for the given task, we estimate that the request is likely to miss its end-to-end SLO since the SLO is divided into latency budgets for individual tasks by the Resource Manager. Therefore, one possible mechanism is to drop the request early on in order to free up resources for requests that are more likely to meet their end-to-end SLOs. Per-task dropping tracks the request at every task during its execution and drops it if it misses the latency budget for any task along the path.
    However, it is important to note that this approach might be overly aggressive, as a request that exceeds its latency budget for an earlier task may still have the potential to meet the end-to-end SLO by compensating at a subsequent task.
    
    \item \textbf{Last-task dropping.} This mechanism does not drop any request up until the last task, even if it exceeds its per-task latency budget at earlier tasks. When the request reaches the last task and its leftover latency budget is smaller than the expected processing time, the request is then dropped. While this approach is more conservative than per-task dropping, it carries the risk of tying up resources at upstream tasks for requests that may ultimately be dropped later.
\end{enumerate}

\textbf{Opportunistic rerouting.}
\label{subsec:opp_rerouting}
To strike a balance between the above-mentioned extreme approaches, we introduce a novel mechanism for early dropping termed \textit{Opportunistic Rerouting}.
This approach involves intelligently redirecting requests that are running behind if there is a chance for them to meet their latency SLOs.

Opportunistic rerouting navigates the tradeoff between being overly aggressive or conservative.
The key idea is that if a request exceeds its latency budget at any given task, we try to find a faster alternative path for the subsequent task in order to make up for it.

We accomplish this as follows. Suppose a request exceeds the latency budget for the given task by $x$ amount of time, indicating the time we need to make up. Once the request completes its execution at the given worker, we identify the downstream worker to forward the request using the routing table, following our standard procedure. Let us denote the profiled execution time of this downstream worker as $y$. To compensate for the $x$ time deficit, we need to find a downstream worker capable of executing within $y-x$ time to offset the exceeded budget for this task. As mentioned in Section~\ref{sec:request_routing}, the Load Balancer propagates a backup table to all workers, listing downstream workers with leftover capacities. We scan this table for a worker whose profiled execution time is at most $y-x$.
If there are multiple such workers, we select the one with the highest accuracy. If there is still a tie, we break it randomly. If no such worker can be found, we drop the request. Note that this entire process takes place in real-time at individual workers during query execution.

Opportunistic rerouting reduces SLO violations by preemptively saving requests from missing their SLOs and trading off accuracy for SLO fulfillment. We compare opportunistic rerouting with the naïve early dropping techniques mentioned above to study its performance benefits in Section~\ref{sec:request_routing_ablation}.


\section{Empirical Evaluation}
\label{sec:evaluation}

We now present our prototype implementation, experimental setup, and our empirical results.

\subsection{Experimental setup}

\textbf{Implementation:}
We implement \projectname{} in $\sim$8K lines of Python code\footnote{Our code is available at \url{https://github.com/UMass-LIDS/Loki}}. We use ONNX runtime \cite{onnxruntime} with the CUDA execution provider to host the host the models on GPUs for efficient inference. We use Gurobi \cite{gurobi} to solve our MILP optimization. Our cluster consists of 20 NVIDIA GTX 1080 Ti GPU workers. 

We extend the discrete-event simulator from ~\cite{ahmad2024proteus} to evaluate our system on a wide range of system parameters. 
This approach aligns with established practices in the field, as DNN inference is known for its high determinism~\cite{gujarati2017swayam, yan2016serf}. Previous works (e.g., \cite{ahmad2024proteus, alpaserve}), typically conduct a core set of experiments on an actual cluster and then compare the results obtained obtained from the cluster with those from simulation to demonstrate the quantitative differences. They then utilize simulation to investigate the impact of a wide range of parameters on the performance of the system. In line with this methodology, we also use our simulator to explore a broad range of parameters and their effects on our system's performance.

\textbf{Pipelines:}
We consider two types of pipelines in our evaluation, both shown in Figure~\ref{fig:inference-pipelines}:

\begin{itemize}
    \item \textit{Traffic analysis.} It first detects the objects in the video frames and then runs fine-tuned car classification or facial recognition models on the detected car and person objects, respectively. We use YOLOv5 \cite{Jocher_YOLOv5_by_Ultralytics_2020} as the object detection model, EfficientNet \cite{tan2019efficientnet} for car classification, and VGG \cite{vggs} for facial recognition.
    \item \textit{Social media.} The social media pipeline detects objects in images and generates suggested captions for the images. It uses ResNet \cite{he2016deep} for image classification and CLIP-ViT \cite{radford2021clip} for image captioning. 
\end{itemize}

We use a total of 32 model variants in our evaluation across the two pipelines.
We normalize the accuracy of each model variant in a model family by the accuracy of its most accurate variant.

\textbf{Datasets:} We use two input datasets. 
\begin{itemize}
    \item \textit{Traffic data.} We use a single day from the Microsoft Azure functions trace \cite{shahrad2020serverless} for query arrivals to drive load for the traffic analysis pipeline. We use shape-preserving transformations to scale the trace in a way that it matches the capacity of our cluster. Since this trace only contains aggregated information of request arrivals but no request content, we use images from the Bellevue traffic dataset \cite{bhardwaj2022ekya} as the request content to perform inference and generate intermediate requests for subsequent tasks.
    \item \textit{Social media.} We use the Twitter trace \cite{twitter_streaming_trace} used by prior inference serving systems \cite{ahmad2024proteus, romero2021infaas} to drive load for the social media pipeline. However, as the Twitter trace also contains only aggregated information about request arrivals but not request content, we use images from the MS-COCO captions dataset \cite{cococaptions} as the content for the requests.
\end{itemize}

\textbf{Evaluation metrics:}
We define metrics to evaluate our system.
\begin{enumerate}
    \item \textit{System accuracy} is the average accuracy experienced by all requests served by the system.
    \item \textit{Cluster utilization} indicates the ratio of workers used at any given time to the total number of workers in the cluster.
    \item \textit{SLO violation ratio} indicates the ratio of requests that miss their SLOs.
\end{enumerate}

Note that a request could miss its SLO in two ways: (i) it finishes past its SLO, (ii) it gets dropped preemptively by the system. In both cases, the system is unable to fulfill the request

\textbf{Baselines for comparison:}
We compare \projectname{}, the first system capable of performing pipeline-aware hardware and accuracy scaling, against two approaches.

\begin{itemize}
    \item \textit{InferLine} \cite{inferline} is a pipeline-aware, but accuracy-agnostic inference serving system. It can perform hardware scaling but requires the clients to specify a single model variant to use for each task in the pipeline and does not support switching between model variants.
    \item \textit{Proteus} \cite{ahmad2024proteus} is an inference serving system that can scale accuracy for single models but is pipeline-agnostic. We set it up to serve inference pipelines by letting it handle each task in the pipeline independently, i.e., it scales accuracy for each task independently since it is unaware of the dependencies between them.
\end{itemize}

\subsection{Performance comparison}

We present an end-to-end comparison of the system performance of Loki against the baselines on the two representative pipelines.

\begin{figure}[t]
    \centering
        \centering
        \includegraphics[width=\linewidth]{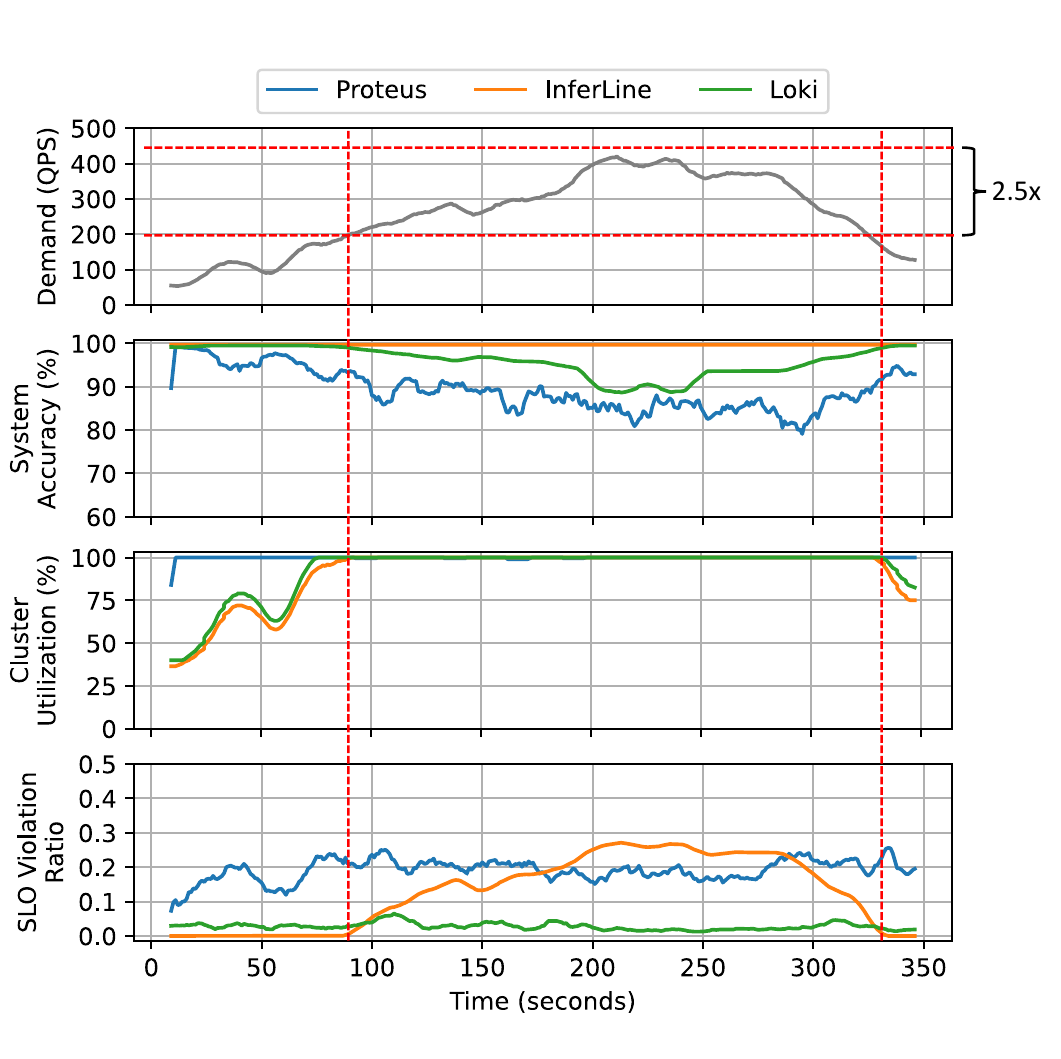}
        \caption{End-to-end comparison on the traffic analysis pipeline. Dotted vertical lines show transition between hardware and accuracy scaling. \projectname{} achieves an increase of $2.5\times$ in effective capacity compared to InferLine that performs hardware scaling alone and reduces SLO violations by up to $10\times$ compared to Proteus that performs pipeline-unaware accuracy scaling.}
    \label{fig:traffic_analysis_endtoend}
\end{figure}

\textbf{Traffic analysis pipeline.} We first study the end-to-end performance of the traffic analysis pipeline.
Figure~\ref{fig:traffic_analysis_endtoend} shows the timeseries of the trace demand, the system accuracy for each approach, the percentage of workers used in the cluster, and the SLO violation ratio. For this experiment, we use an end-to-end pipeline latency SLO of 250ms. We explore the sensitivity of the system to different SLO values in Section~\ref{sec:slo_variation}.

We show the point when \projectname{} shifts between hardware scaling and accuracy scaling with the help of the dotted vertical lines. InferLine offers low SLO violations during the hardware scaling phase, but since it is not capable of performing accuracy scaling, its SLO violations shoot up during that time and it is not able to meet the increased demand. Therefore, compared to InferLine which performs hardware scaling alone, \projectname{} effectively increases the capacity of the cluster by $2.5\times$.

Proteus consistently suffers from high SLO violations due to the fact that it is not pipeline-aware and manages each task in the pipeline graph independently. Therefore, Proteus is not able to identify the dependencies between the tasks to match the throughput of different tasks, leading to the creation of throughput bottlenecks. Therefore, \projectname{} reduces SLO violations by $10\times$ compared to a pipeline-unaware accuracy scaling approach such as Proteus.

As \projectname{} performs accuracy scaling in a pipeline-aware manner, it is also able to achieve higher system accuracy than Proteus since the latter may drop accuracy for a task that may lead to a higher drop in end-to-end accuracy, while \projectname{} uses the knowledge of end-to-end accuracies to drop minimal accuracy.

Lastly, during off-peak times, \projectname{} can leverage hardware scaling to reduce cost and energy by allowing the system to shut down servers that are not needed. Compared to Proteus which uses the entire cluster throughout since it does not perform hardware scaling, \projectname{} reduces the number of servers needed to serve the demand, and consequently server cost, by up to $2.67\times$.

To summarize, \projectname{} offers consistently lower SLO violations due to its pipeline-aware resource allocation. It increases the effective capacity of the cluster by $2.5\times$ in this experiment, and can shut off servers to save cost and energy during off-peak times.

\textbf{Validating the simulator.} We conduct this experiment on our simulator as well to validate it and observe an average difference of 1.2\% in accuracy, 1.8\% in the SLO violation ratios, and 1.5\% in the number of servers used. We note that the simulation results are close to the prototype results, and the differences are produced due to various factors such as small variances in model execution times and unexpected network delays. However, due to the deterministic nature of ML inference and this small difference, we use our simulator to conduct the remaining experiments in order to evaluate the system under a wide range of conditions and parameters. For the rest of this Section, we present results from our simulation unless otherwise noted.

\begin{figure}[t]
    \centering
        \centering
        \includegraphics[width=\linewidth]{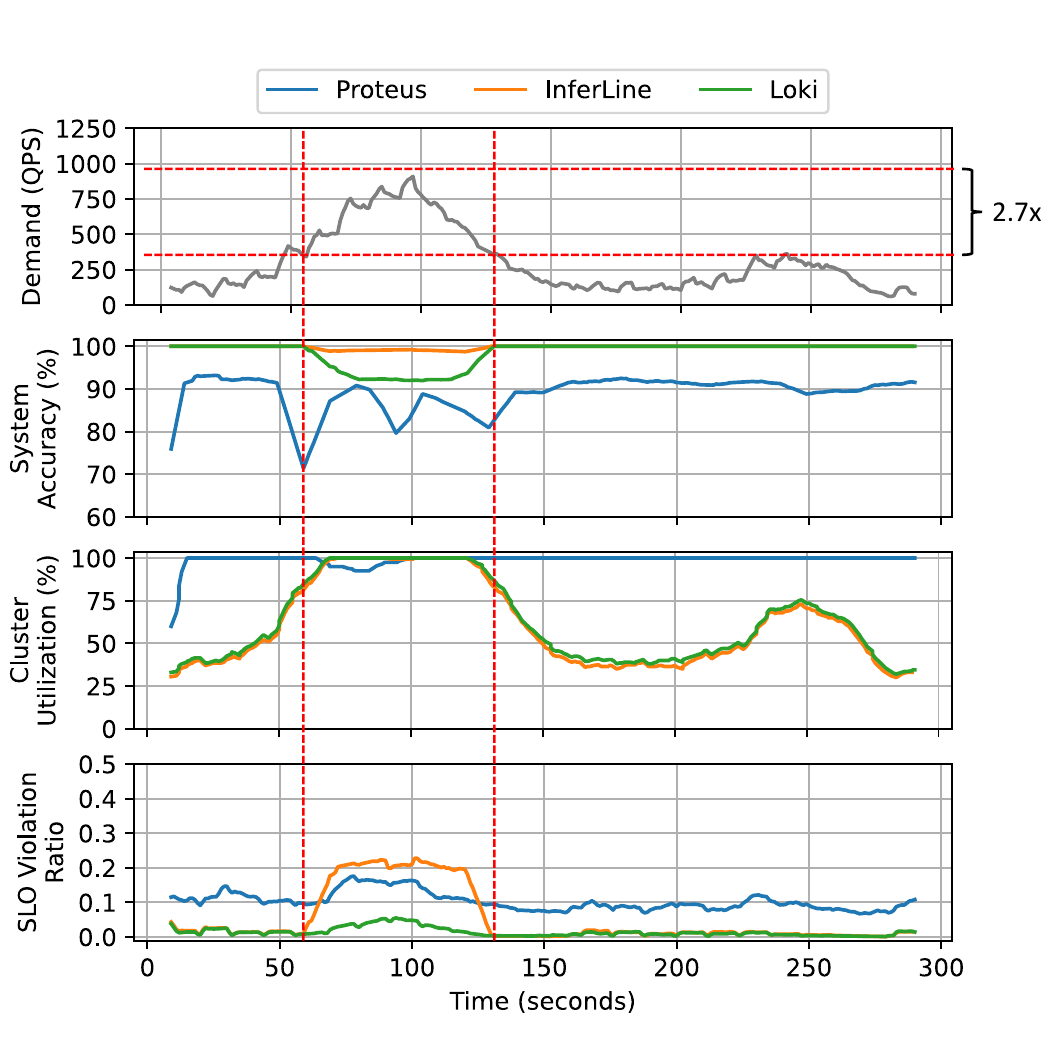}
        \caption{End-to-end comparison on the social media pipeline. Dotted vertical lines show transition between hardware and accuracy scaling. \projectname{} achieves an increase of $2.7\times$ in effective capacity compared to hardware scaling alone and reduces SLO violations by up to $10\times$ compared to pipeline-unaware accuracy scaling.}
    \label{fig:social_media_endtoend}
\end{figure}

\textbf{Social media pipeline.} We now present the end-to-end performance comparison on the social media pipeline in Figure~\ref{fig:social_media_endtoend}. As before, we show the incoming demand into the system, the system accuracy offered by the different approaches, cluster utilization, and SLO violation ratio.

We observe similar trends as in the traffic analysis pipeline. When demand increases to the point where hardware scaling is not able to meet it, the SLO violations of InferLine shoot up to more than $5\times$ of \projectname{}. During this time, \projectname{} is able to meet demand by sacrificing $\sim$10\% accuracy.

During off-peak times, \projectname{} again uses about $2.67\times$ less servers than Proteus which does not perform any hardware scaling. \projectname{} also drops up to $20\%$ less accuracy than Proteus due to the ability of the former to identify pipeline dependencies and their effect on end-to-end accuracy of the pipeline. Proteus continues to suffer from high violations again due to being pipeline-unaware.
\projectname{} increases the effective capacity of the cluster by $2.7\times$ in this experiment.

\subsection{Ablation study of the load balancer}
\label{sec:request_routing_ablation}

\begin{figure}[t]
    \centering
        \centering
        \includegraphics[width=0.7\linewidth]{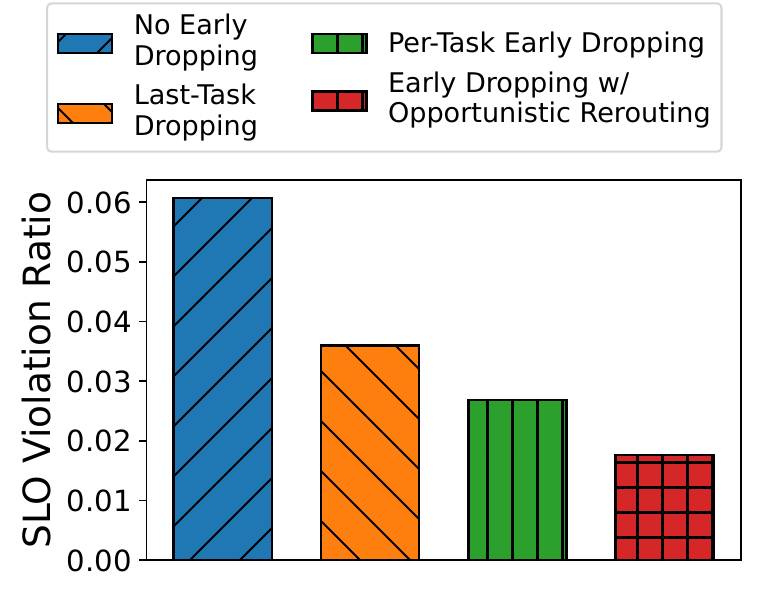}
        \caption{Ablation study of the load balancer shows that opportunistic rerouting has the most impact on SLO violations.}
    \label{fig:request_routing_ablation}
\end{figure}

We now take a deep dive into the request routing performed by the Load Balancer to understand where the performance benefits come from.
Figure~\ref{fig:request_routing_ablation} shows the benefit achieved from the use of early dropping and opportunistic rerouting by comparing it against simpler versions as follows.

\begin{enumerate}
    \item \textit{No early dropping:} This is the simplest version which does not perform any early dropping and follows the original routing plan.
    \item \textit{Last-task dropping:} This version drops requests if they are expected to miss their SLOs, but only at the last task of the pipeline.
    \item \textit{Per-task early dropping:} This version performs early dropping of requests at each task if they miss the assigned latency budget of that task.
    \item \textit{Early dropping with opportunistic rerouting:} This is the full-fledged version of our approach that we use in our end-to-end implementation. It first tries to re-route requests through faster paths if they are expected to miss their SLO using the assigned path, and drops them if this is not possible.
\end{enumerate}

We observe that the version without any early dropping suffers from the highest SLO violations as it can waste resources on requests that are not on target to meet their SLOs, hence delaying and potentially timing out other requests as well. Last-task dropping improves SLO violations slightly by dropping requests if they are expected to miss SLOs, but since it only does this at the last task, it can be overly conservative in doing so and still suffers from high SLO violations. We observe that per-task early dropping improves performance further by dropping requests at each task if they are expected to miss the latency budget for the respective task. However, this approach may drop requests too aggressively since a request that misses its latency budget for an earlier task may still potentially catch up at a later task.

Our approach, which opportunistically reroutes requests that are falling back to faster paths, minimizes SLO violations the most. If such rerouting is not possible, it means that the request has no way of meeting its SLO even if routed through the fastest path available. In this case, it drops the requests as a last resort in order to free up resources for other requests that may have a better chance of meeting their SLOs.

\subsection{Effect of SLOs on system performance}
\label{sec:slo_variation}

\begin{figure}[t]
    \centering
        \centering
        \includegraphics[width=\linewidth]{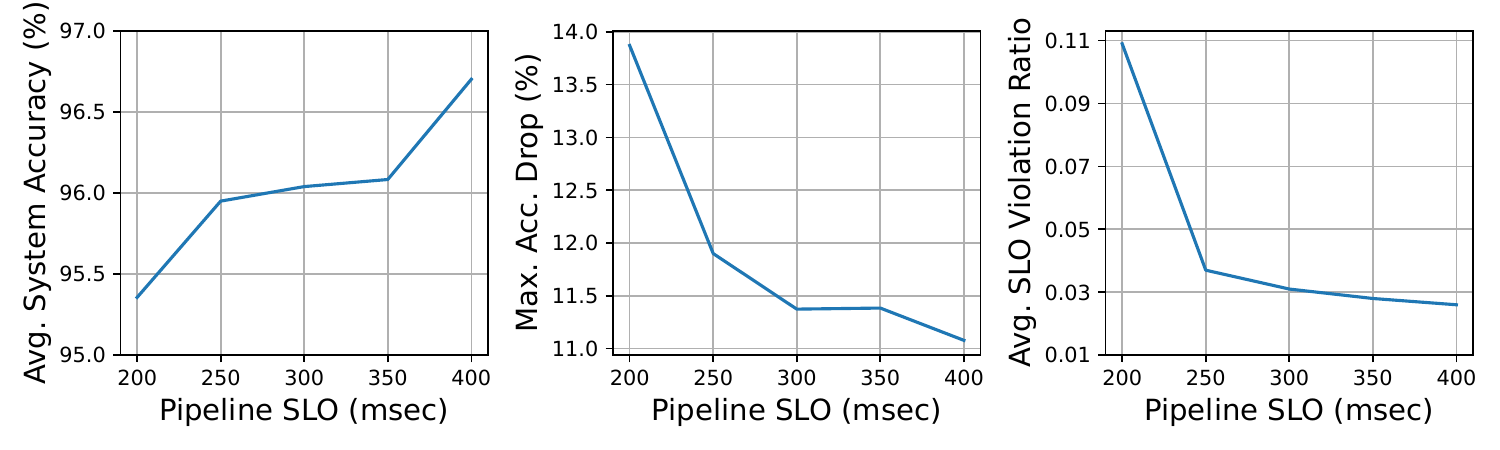}
        \caption{Effect of varying SLOs on \projectname{}
        }
    \label{fig:effect_of_slo}
\end{figure}

We study the effect of varying the latency SLOs for the traffic analysis pipeline on the performance of Loki. To summarize the results from a large number of experiments, Figure~\ref{fig:effect_of_slo} shows the following key metrics: (i) the average system accuracy across the entire experiment, (ii) the maximum accuracy drop, and (iii) the average SLO violation ratio. The maximum accuracy drop is the degradation in system accuracy from its highest possible value at peak demand.

We observe a general trend that performance improves sharply with initial increments of 50 milliseconds, but there are diminishing improvements in performance as we use larger values of SLO. This is because the optimization can use several knobs to meet tighter SLOs: (i) creating more replicas of model instances, (ii) decreasing the batch size of models in the path to lower end-to-end latency, and (iii) lowering accuracy by changing the model variant. Note that the Resource Manager can only increase the replication factor up to the point where the entire cluster is allocated, and the minimum batch size it can use is 1. Starting from 400 milliseconds, as the latency SLO gets tighter, the system can first respond by using these knobs without sacrificing any accuracy. However, when the system faces even tighter latency SLOs, once it exhausts these options, the system is compelled to resort to accuracy scaling to meet SLOs. This results in a decrease in overall system accuracy and leads to SLO violations due to the overhead associated with swapping model variants.

Below 200 milliseconds, the system cannot serve the demand even with the maximum degree of hardware and accuracy scaling because the sum of processing latencies across the entire pipeline of even the lowest accuracy model variants with a batch size of 1 exceeds this value of SLO.

\subsection{Runtime performance}
\label{sec:milp_overhead}

We now explore the runtime performance of both the core components of our system: the Resource Manager and Load Balancer.

\textbf{Resource Manager.} Given that the Resource Manager considers all paths through the pipeline and yields an optimal solution by solving an MILP, it is expected to run orders of magnitude slower than the Load Balancer. We measure the average runtime of the MILP to be $\sim$500 milliseconds. 
As the Resource Manager is invoked periodically to adapt to long-term fluctuations in demand and does not lie on the critical path of query execution, the observed runtime allows for a reasonably swift adaptation of resource allocation in response to changing demands.

\textbf{Load Balancer} As the Load Balancer reacts to run-time changes in demand, it needs to respond much faster than the Resource Manager. In our experiments, we measure the average runtime of the load balancer to be $\sim$0.15 milliseconds. We attribute the fast runtime of the Load Balancer to the efficiency of our request routing algorithm presented in Section~\ref{sec:request_routing}.



\section{Related Work}
\label{sec:related}
Inference serving is quickly becoming a hot topic of research.
Representative production systems include TensorFlow-Serving \cite{tfserving}, NVIDIA Triton Inference Server \cite{triton} and Amazon SageMaker \cite{sagemaker}. Inference serving has also been extensively studied through research prototypes as well, such as Clipper \cite{clipper}, INFless \cite{yang2022infless}, and PRETZEL \cite{lee2018pretzel}.
These systems aim to provide a unified abstraction to the user to hide details of the underlying ML frameworks, data pre-processing, and performance optimization. Unlike these systems that require users to manage DNN models, \projectname{} automatically configures the suitable DNN models to execute on GPU clusters.

The closest works to \projectname{} are Proteus \cite{ahmad2024proteus} and InferLine \cite{inferline}. Proteus presents an inference serving system that can scale accuracy for single models but is pipeline-agnostic. It scales accuracy for each task in the pipeline independently since it is unaware of the dependencies between them. InferLine is a pipeline-aware inference serving system that minimizes the cost of inference serving by scaling the hardware in response to changes in demand.

INFaaS \cite{romero2021infaas}, Sommelier~\cite{guo2022sommelier}, and Tolerance Tiers~\cite{halpern2019one} are also related as they also consider model variants with different accuracy-latency profiles in model serving systems.
INFaaS~\cite{romero2021infaas} presents a model-less inference serving system that automates the selection of model variants for each query to minimize cost while meeting accuracy and latency requirements. 
Unlike \projectname{} that explicitly optimizes accuracy as an objective, it treats accuracy as a constraint and focuses on hardware scaling to handle variable demands. 
Sommelier \cite{guo2022sommelier} is a model repository that can interface with inference serving systems to suggest model variants with lower accuracy to handle increases in load. Tolerance Tiers~\cite{halpern2019one} allows developers to tradeoff accuracy for latency through programming APIs. However, it imposes a fundamental limitation on applications, compelling them to adhere to a single accuracy tier statically throughout the entire inference serving process, lacking the flexibility to dynamically adjust accuracy as part of a scaling approach.

Many inference serving systems try to optimize the cost of serving while meeting certain performance constraints. Kairos \cite{kairos} is one such system that aims to minimize the cost of inference serving using heterogeneous cloud resources. MArk \cite{zhang2019mark} and Scrooge \cite{scrooge} also try to minimize the cost of inference serving while trying to meet latency SLOs. iGniter \cite{xu2023igniter} is an interference-aware inference serving system that minimizes serving cost. Cocktail \cite{gunasekaran2022cocktail} uses model ensembling to improve accuracy and meet latency requirements using minimal cost. \projectname{} instead optimizes both cost and accuracy by unifying accuracy scaling and hardware scaling.

Some model serving systems propose techniques that can be combined with accuracy and hardware scaling to improve system throughput.  
Rafiki \cite{wang2018rafiki} is an analytics serving system that uses model ensembling during inference to improve accuracy at the cost of latency.
PERSEUS \cite{lemay2020perseus} studies the performance and cost tradeoffs associated with multi-tenant model serving. 
Morphling \cite{wang2021morphling} presents an algorithmic framework to minimize the cost of searching through possible configurations when setting up inference services.
Clover \cite{li2023clover} is an inference serving system that explores the tradeoff between carbon emissions and accuracy. 
DeepPlan \cite{jeong2023deepplan} minimizes inference latency by exploiting recent advances in GPU technology to reduce the model loading latencies.
SHEPHERD \cite{zhang2023shepherd} and Clockwork \cite{gujarati2020serving} aim to minimize the tail latency of model serving by eliminating sources of unpredictability in the system.

There has been a lot of work specifically related to video analytics pipelines. VideoStorm \cite{zhang2017videostorm} was the first work to explore the latency-accuracy tradeoff for the resource provisioning of video analytics applications that use DNNs. Llama \cite{llama2021romero} is a serverless framework for auto-tuning video analytics pipelines. Nexus \cite{shen2019nexus} is another framework for serving video analytics pipelines on GPU clusters. In comparison, \projectname{} is a system that is applicable to generic inference pipelines that can be represented as directed rooted trees (defined in Section~\ref{sec:background}). 

Recent work also explores serving large language models (LLMs), such as AlpaServe \cite{alpaserve} and Tabi \cite{wang2023tabi}. LLM serving is different from traditional inference serving in the sense that it often requires partitioning the model to be served by multiple servers. \projectname{} does not feature optimizations tailored to LLMs but can cater to inference pipelines with LLMs. 

\section{Conclusion}
\label{sec:conclusion}
In conclusion, our work addresses the pressing need for efficient and cost-effective deployment of machine learning (ML) inference at the edge. By recognizing the challenge posed by limited edge resources and the computational demands of ML models, we introduce \projectname{}, a system for resource provisioning of ML inference pipelines. Central to Loki is the concept of hardware and accuracy scaling, which dynamically adjusts accuracy levels to manage resource constraints when needed, thereby enhancing the effective capacity of edge clusters and minimizing resource usage during the off-peak. Our experimental results demonstrate that Loki significantly outperforms existing inference serving systems by reducing Service Level Objective (SLO) violations by up to $10\times$ and increasing the effective capacity by up to $2.7\times$ while sacrificing minimal accuracy and meeting throughput targets.


\begin{acks}
This work is supported in part by the National Science Foundation under grants CNS-1763617, CNS-1901137, CNS-2106463, CNS-2312396, CNS-2338512, CNS-2224054, and DMS-2220211.
\end{acks}

\bibliographystyle{ACM-Reference-Format}
\bibliography{references}

\end{document}